\pdfoutput=1
\documentclass[12pt,a4paper,titlepage]{article}
\usepackage[utf8]{inputenc}
\usepackage[top=50pt,bottom=50pt,left=68pt,right=66pt]{geometry}
\usepackage{amsmath}
\usepackage{booktabs}
\usepackage{amssymb}
\usepackage[title]{appendix}
\usepackage{mathrsfs}
\usepackage{float}
\usepackage{multirow}
\usepackage{graphicx,caption,subcaption}
\usepackage[space]{grffile}

\begin{document}
\renewcommand{\arraystretch}{1.3}

\makeatletter
\def\@hangfrom#1{\setbox\@tempboxa\hbox{{#1}}%
      \hangindent 0pt
      \noindent\box\@tempboxa}
\makeatother


\def\un#1{\relax\ifmmode\@@underline#1\else
        $\@@underline{\hbox{#1}}$\relax\fi}


\let\under=\unt                 
\let\ced=\ce                    
\let\du=\du                     
\let\um=\Hu                     
\let\sll=\lp                    
\let\Sll=\Lp                    
\let\slo=\os                    
\let\Slo=\Os                    
\let\tie=\ta                    
\let\br=\ub                     


\def\a{\alpha}
\def\b{\beta}
\def\c{\chi}
\def\d{\delta}
\def\e{\epsilon}
\def\f{\phi}
\def\g{\gamma}
\def\h{\eta}
\def\i{\iota}
\def\j{\psi}
\def\k{\kappa}
\def\l{\lambda}
\def\m{\mu}
\def\n{\nu}
\def\o{\omega}
\def\p{\pi}
\def\q{\theta}
\def\r{\rho}
\def\s{\sigma}
\def\t{\tau}
\def\u{\upsilon}
\def\x{\xi}
\def\z{\zeta}
\def\D{\Delta}
\def\F{\Phi}
\def\G{\Gamma}
\def\J{\Psi}
\def\L{\Lambda}
\def\O{\Omega}
\def\P{\Pi}
\def\Q{\Theta}
\def\S{\Sigma}
\def\U{\Upsilon}
\def\X{\Xi}


\def\ve{\varepsilon}
\def\vf{\varphi}
\def\vr{\varrho}
\def\vs{\varsigma}
\def\vq{\vartheta}


\def\ca{{\cal A}}
\def\cb{{\cal B}}
\def\cc{{\cal C}}
\def\cd{{\cal D}}
\def\ce{{\cal E}}
\def\cf{{\cal F}}
\def\cg{{\cal G}}
\def\ch{{\cal H}}
\def\ci{{\cal I}}
\def\cj{{\cal J}}
\def\ck{{\cal K}}
\def\cl{{\cal L}}
\def\cm{{\cal M}}
\def\cn{{\cal N}}
\def\co{{\cal O}}
\def\cp{{\cal P}}
\def\cq{{\cal Q}}
\def\car{{\cal R}}
\def\cs{{\cal S}}
\def\ct{{\cal T}}
\def\cu{{\cal U}}
\def\cv{{\cal V}}
\def\cw{{\cal W}}
\def\cx{{\cal X}}
\def\cy{{\cal Y}}
\def\cz{{\cal Z}}


\def\Sc#1{{\hbox{\sc #1}}}      
\def\Sf#1{{\hbox{\sf #1}}}      



\def\slpa{\slash{\pa}}                            
\def\slin{\SLLash{\in}}                                   
\def\bo{{\raise-.3ex\hbox{\large$\Box$}}}               
\def\cbo{\Sc [}                                         
\def\pa{\partial}                                       
\def\de{\nabla}                                         
\def\dell{\bigtriangledown}                             
\def\su{\sum}                                           
\def\pr{\prod}                                          
\def\iff{\leftrightarrow}                               
\def\conj{{\hbox{\large *}}}                            
\def\ltap{\raisebox{-.4ex}{\rlap{$\sim$}} \raisebox{.4ex}{$<$}}   
\def\gtap{\raisebox{-.4ex}{\rlap{$\sim$}} \raisebox{.4ex}{$>$}}   
\def\TH{{\raise.2ex\hbox{$\displaystyle \bigodot$}\mskip-4.7mu \llap H \;}}
\def\face{{\raise.2ex\hbox{$\displaystyle \bigodot$}\mskip-2.2mu \llap {$\ddot
        \smile$}}}                                      
\def\dg{\sp\dagger}                                     
\def\ddg{\sp\ddagger}                                   

\font\tenex=cmex10 scaled 1200


\def\sp#1{{}^{#1}}                              
\def\sb#1{{}_{#1}}                              
\def\oldsl#1{\rlap/#1}                          
\def\slash#1{\rlap{\hbox{$\mskip 1 mu /$}}#1}      
\def\Slash#1{\rlap{\hbox{$\mskip 3 mu /$}}#1}      
\def\SLash#1{\rlap{\hbox{$\mskip 4.5 mu /$}}#1}    
\def\SLLash#1{\rlap{\hbox{$\mskip 6 mu /$}}#1}      
\def\PMMM#1{\rlap{\hbox{$\mskip 2 mu | $}}#1}   %
\def\PMM#1{\rlap{\hbox{$\mskip 4 mu ~ \mid $}}#1}       %
\def\Tilde#1{\widetilde{#1}}                    
\def\Hat#1{\widehat{#1}}                        
\def\Bar#1{\overline{#1}}                       
\def\sbar#1{\stackrel{*}{\Bar{#1}}}             
\def\bra#1{\left\langle #1\right|}              
\def\ket#1{\left| #1\right\rangle}              
\def\VEV#1{\left\langle #1\right\rangle}        
\def\abs#1{\left| #1\right|}                    
\def\leftrightarrowfill{$\mathsurround=0pt \mathord\leftarrow \mkern-6mu
        \cleaders\hbox{$\mkern-2mu \mathord- \mkern-2mu$}\hfill
        \mkern-6mu \mathord\rightarrow$}
\def\dvec#1{\vbox{\ialign{##\crcr
        \leftrightarrowfill\crcr\noalign{\kern-1pt\nointerlineskip}
        $\hfil\displaystyle{#1}\hfil$\crcr}}}           
\def\dt#1{{\buildrel {\hbox{\LARGE .}} \over {#1}}}     
\def\dtt#1{{\buildrel \bullet \over {#1}}}              
\def\der#1{{\pa \over \pa {#1}}}                
\def\fder#1{{\d \over \d {#1}}}                 


\def\frac#1#2{{\textstyle{#1\over\vphantom2\smash{\raise.20ex
        \hbox{$\scriptstyle{#2}$}}}}}                   
\def\half{\frac12}                                        
\def\sfrac#1#2{{\vphantom1\smash{\lower.5ex\hbox{\small$#1$}}\over
        \vphantom1\smash{\raise.4ex\hbox{\small$#2$}}}} 
\def\bfrac#1#2{{\vphantom1\smash{\lower.5ex\hbox{$#1$}}\over
        \vphantom1\smash{\raise.3ex\hbox{$#2$}}}}       
\def\afrac#1#2{{\vphantom1\smash{\lower.5ex\hbox{$#1$}}\over#2}}    
\def\partder#1#2{{\partial #1\over\partial #2}}   
\def\parvar#1#2{{\d #1\over \d #2}}               
\def\secder#1#2#3{{\partial^2 #1\over\partial #2 \partial #3}}  
\def\on#1#2{\mathop{\null#2}\limits^{#1}}               
\def\bvec#1{\on\leftarrow{#1}}                  
\def\oover#1{\on\circ{#1}}                              

\def\[{\lfloor{\hskip 0.35pt}\!\!\!\lceil}
\def\]{\rfloor{\hskip 0.35pt}\!\!\!\rceil}
\def\Lag{{\cal L}}
\def\du#1#2{_{#1}{}^{#2}}
\def\ud#1#2{^{#1}{}_{#2}}
\def\dud#1#2#3{_{#1}{}^{#2}{}_{#3}}
\def\udu#1#2#3{^{#1}{}_{#2}{}^{#3}}
\def\calD{{\cal D}}
\def\calM{{\cal M}}

\def\szet{{${\scriptstyle \b}$}}
\def\ulA{{\un A}}
\def\ulM{{\underline M}}
\def\cdm{{\Sc D}_{--}}
\def\cdp{{\Sc D}_{++}}
\def\vTheta{\check\Theta}
\def\fracm#1#2{\hbox{\large{${\frac{{#1}}{{#2}}}$}}}
\def\ha{{\fracmm12}}
\def\tr{{\rm tr}}
\def\Tr{{\rm Tr}}
\def\itrema{$\ddot{\scriptstyle 1}$}
\def\ula{{\underline a}} \def\ulb{{\underline b}} \def\ulc{{\underline c}}
\def\uld{{\underline d}} \def\ule{{\underline e}} \def\ulf{{\underline f}}
\def\ulg{{\underline g}}
\def\items#1{\\ \item{[#1]}}
\def\ul{\underline}
\def\un{\underline}
\def\fracmm#1#2{{{#1}\over{#2}}}
\def\footnotew#1{\footnote{\hsize=6.5in {#1}}}
\def\low#1{{\raise -3pt\hbox{${\hskip 0.75pt}\!_{#1}$}}}

\def\Dot#1{\buildrel{_{_{\hskip 0.01in}\bullet}}\over{#1}}
\def\dt#1{\Dot{#1}}

\def\DDot#1{\buildrel{_{_{\hskip 0.01in}\bullet\bullet}}\over{#1}}
\def\ddt#1{\DDot{#1}}

\def\DDDot#1{\buildrel{_{_{\hskip 0.01in}\bullet\bullet\bullet}}\over{#1}}
\def\dddt#1{\DDDot{#1}}

\def\DDDDot#1{\buildrel{_{_{\hskip 
0.01in}\bullet\bullet\bullet\bullet}}\over{#1}}
\def\ddddt#1{\DDDDot{#1}}

\def\Tilde#1{{\widetilde{#1}}\hskip 0.015in}
\def\Hat#1{\widehat{#1}}


\newskip\humongous \humongous=0pt plus 1000pt minus 1000pt
\def\caja{\mathsurround=0pt}
\def\eqalign#1{\,\vcenter{\openup2\jot \caja
        \ialign{\strut \hfil$\displaystyle{##}$&$
        \displaystyle{{}##}$\hfil\crcr#1\crcr}}\,}
\newif\ifdtup
\def\panorama{\global\dtuptrue \openup2\jot \caja
        \everycr{\noalign{\ifdtup \global\dtupfalse
        \vskip-\lineskiplimit \vskip\normallineskiplimit
        \else \penalty\interdisplaylinepenalty \fi}}}
\def\li#1{\panorama \tabskip=\humongous                         
        \halign to\displaywidth{\hfil$\displaystyle{##}$
        \tabskip=0pt&$\displaystyle{{}##}$\hfil
        \tabskip=\humongous&\llap{$##$}\tabskip=0pt
        \crcr#1\crcr}}
\def\eqalignnotwo#1{\panorama \tabskip=\humongous
        \halign to\displaywidth{\hfil$\displaystyle{##}$
        \tabskip=0pt&$\displaystyle{{}##}$
        \tabskip=0pt&$\displaystyle{{}##}$\hfil
        \tabskip=\humongous&\llap{$##$}\tabskip=0pt
        \crcr#1\crcr}}


\def\eV{\,{\rm eV}}
\def\keV{\,{\rm keV}}
\def\MeV{\,{\rm MeV}}
\def\GeV{\,{\rm GeV}}
\def\TeV{\,{\rm TeV}}
\def\sv{\left<\sigma v\right>}
\def\({\left(}
\def\){\right)}
\def\cm{{\,\rm cm}}
\def\K{{\,\rm K}}
\def\kpc{{\,\rm kpc}}
\def\beq{\begin{equation}}
\def\eeq{\end{equation}}
\def\bea{\begin{eqnarray}}
\def\eea{\end{eqnarray}}


\newcommand{\be}{\begin{equation}}
\newcommand{\ee}{\end{equation}}
\newcommand{\nbe}{\begin{equation*}}
\newcommand{\nee}{\end{equation*}}

\newcommand{\fr}{\frac}
\newcommand{\lb}{\label}

\thispagestyle{empty}

{\hbox to\hsize{
\vbox{\noindent October 2022 \hfill IPMU22-0031} }}

\noindent  ~revised version 

\noindent
\vskip2.0cm
\begin{center}

{\large\bf Starobinsky-Bel-Robinson gravity}

\vglue.4in

Sergei V. Ketov~${}^{a,b,c,\#}$ 
\vglue.3in

${}^a$~Department of Physics, Tokyo Metropolitan University,\\
1-1 Minami-ohsawa, Hachioji-shi, Tokyo 192-0397, Japan \\
${}^b$~Tomsk State University, 36 Lenin Avenue, Tomsk 634050, Russia\\
${}^c$~Kavli Institute for the Physics and Mathematics of the Universe (WPI),
\\The University of Tokyo Institutes for Advanced Study,  Kashiwa 277-8583, Japan\\
\vglue.1in

${}^{\#}$~ketov@tmu.ac.jp
\end{center}

\vglue.4in

\begin{center}
{\Large\bf Abstract}  
\end{center}

A novel superstring-inspired gravitational theory in four spacetime dimensions is proposed as a sum of the modified $(R+\a R^2)$ gravity motivated by the Starobinsky inflation and the Bel-Robinson-tensor-squared term motivated by the eleven-dimensional 
M-theory dimensionally reduced to four dimensions. The proposed Starobinsky-Bel-Robinson action has only two  parameters, which makes it suitable for verifiable physical applications in black hole physics, cosmological inflation and Hawking radiation in the early universe.

\newpage

\section{Introduction}

General relativity theory with the Einstein-Hilbert (EH) action for gravity in $D=4$ spacetime dimensions is well confirmed by precision measurements inside the Solar system. However, the EH action has to be modified in the UV-regime (for high energies and curvatures, in early Universe), in the IR-regime (for cosmological distances), and (beyond any doubt)  in quantum gravity. When preserving the general coordinate invariance and locality, the EH gravity action can only be modified (besides a cosmological constant) by extra terms of the higher order in spacetime curvature. Those higher-derivative terms in the gravitational effective action are supposed to describe quantum gravity corrections to the EH action, while they are necessary for physical applications in the UV-regime because the EH gravity is non-renormalizable. This issue is well known in the literature about gravity but the main problem is a well motivated derivation of the gravitational effective action from quantum gravity or a reasonable selection of the higher order terms because there are infinitely many of them. 

Solving this problem requires a practical framework for quantum gravity in order to perform calculations, which is another problem. Superstring theory is a mathematically consistent framework for quantum gravity but its applications to observed physics are limited by the necessity of compactification of extra  $(D-4)$ spacetime dimensions and related huge uncertainties in verifiable predictions. Moreover, superstring theory in $D=10$ dimensions is defined as a quantum perturbation theory and is not background-independent. Actually, string theory can only be formulated on Ricci-flat backgrounds and does not allow de Sitter vacua.

Nevertheless, it makes sense to use insights from superstrings/M-theory together with other insights into quantum gravity, coming from early universe cosmology, black hole physics  and particle physics beyond the Standard Model, in order to motivate the {\it  leading} quantum gravity corrections to the EH action of gravity.

In this letter we give a new proposal for some leading quantum corrections in the high-curvature regime, which are motivated by M-theory in $D=11$ dimensions after its dimensional reduction to $D=4$ dimensions, combined with the simplest viable model of cosmological inflation in four space-time dimensions, known as the Starobinsky model. The resulting modified gravity action is called the Starobinsky-Bel-Robinson (SBR) gravity.

Our paper is organized as follows. In Sec.~2 we recall the Starobinsky model of inflation in four spacetime dimensions and its possible origin from higher dimensions. In Sec.~3 we recall the low-energy effective action of M-theory in  eleven dimensions and its possible contribution to the gravitational effective action in four spacetime dimensions. In Sec.~4 
we formulate the novel four-dimensional SBR gravity and give our conclusion.

\section{Starobinsky gravity and extra dimensions}

The Starobinsky model of inflation is defined by the modified gravity action \cite{Starobinsky:1980te}
 \be \label{star}
S_{\rm Star.} = \fracmm{M^2_{\rm Pl}}{2}\int \mathrm{d}^4x\sqrt{-g} \left( R +\fracmm{1}{6m^2}R^2\right)~,
\ee
where we have introduced the scalar curvature $R$, reduced Planck mass $M_{\rm Pl}=1/\sqrt{8\p G_{\rm N}}\approx 2.4\times 10^{18}$ GeV, and the mass parameter $m$. We use the spacetime signature $(-,+,+,+,)$. 

In the low curvature spacetime, the $R^2$ term can be ignored and the action (\ref{star}) reduces to the EH action. During inflation in the early Universe (with strong spacetime curvature), the EH-term can be ignored and the action  (\ref{star}) reduces to the no-scale $R^2$ gravity with the dimensionless coupling constant in front of the action. The $R^2$-term with a positive coupling constant is the {\it only\/} term in a generic Lagrangian, quadratic in the curvature tensor,  that does not lead to ghosts, i.e. it is well theoretically motivated. The quantized gravity theory (\ref{star}) is, however, non-renormalizable, with the UV cutoff being given by $M_{\rm Pl}$, see e.g., Ref.~\cite{Ketov:2022qwj} for more details.

The action ({\ref{star}) is also well phenomenologically motivated due to its excellent agreement with
WMAP/Planck/BICEP/KECK precision measurements of the cosmic microwave background radiation \cite{BICEP:2021xfz}. Then the parameter $m$ is the inflaton mass that can be fixed by the COBE/WMAP normalization as
\be \lb{starm}
m\approx 3 \cdot10^{13}~{\rm GeV} \quad {\rm or}\quad \fracmm{m}{M_{\rm Pl}}\approx 1.3\cdot 10^{-5}~.
\ee

The action (\ref{star}) can be deduced from higher $(D)$ spacetime dimensions, while preserving the hierarchy of physical scales, $H_{\rm inf.} \ll M_{\rm KK} \ll M_{\rm Pl}$,
where $H_{\rm inf.}$ is the Hubble scale of inflation, and $M_{\rm KK}$ is the Kaluza-Klein scale.  The relevant field theory in $D>4$ dimensions should have the modified gravity Lagrangian of the form $(R+\a R^n)$ in terms of the $D$-dimensional scalar curvature $R$, coupled to a $(p-1)$-form gauge field $A$ and a having cosmological constant. Then the warped compactification from $D$ dimensions on a sphere $S^{D-4}$ with a non-vanishing flux of the gauge field strength $F=d\wedge A$ down to four space-time dimensions leads to  the Starobinsky action (\ref{star}). As was demonstrated in Refs.~\cite{Otero:2017thw,Nakada:2017uka}, it is only possible when $n=D/2$ and $p=n$. In particular, when $D=8$, the modulus (radius) of the hidden four-sphere $S^4$ of extra dimensions can be stabilized, while the modified $D=8$ gravity can be embedded into the modified $D=8$ (Salam-Sezgin) gauged supergravity  \cite{Salam:1984ft}. In turn, as was argued in Ref.~\cite{Nakada:2017uka}, the modified $D=8$ supergravity may be derived by compactifying the modified 11-dimensional supergravity on a three-sphere $S^3$. As a result, the Starobinsky action (\ref{star}) may be derivable from the 11-dimensional supergravity modified by the term quartic in the scalar curvature and compactified on the product $S^4\times S^3$ of extra dimensions. Unfortunately, a supersymmetric completion of such action in $D=11$ was never
found. The significance of eleven dimensions is due to the fact that the supergravity theory in $D=11$ is {\it unique} \cite{Cremmer:1978km}, while its extension to quantum gravity, known under the name of M-theory, is presumably also unique, with local supersymmetry in $D=11$ playing the essential role.~\footnote{Another possibility is to start from F-theory in $D=12$ dimensions and compactify it on a product of two Kummer surfaces $K3\times K3$.}

\section{Bel-Robinson tensor squared term}

Having recognized the importance of eleven spacetime dimensions, it is natural to ask what could be the
next term in the effective gravity Lagrangian beyond the $R^2$ term in $D=4$, when starting from M-theory beyond the supergravity action in $D=11$, after its (flux) compactification to four spacetime dimensions.

The bosonic terms (when all fermionic fields are ignored) of M-theory in the leading order beyond the  $D=11$ supergravity action read \cite{Green:1997di,Tseytlin:2000sf}
\begin{align} \label{m11}
S_{\rm M} & =  \fracmm{1}{2\k^2_{11}}\int d^{11}x\,\sqrt{-g}\left[ R -
\fracmm{1}{2\cdot 4!}F^2 -\fracmm{1}{6\cdot 3!\cdot(4!)^2}\ve_{11}CFF\right] \\
\nonumber
&  -\fracmm{T_2}{(2\p)^4\cdot 3^2\cdot 2^{13}}\int d^{11}x\,\sqrt{-g}
\left( J_{11} - \fracmm{1}{2} E_8 \right) + T_2 \int C\wedge X_8~~,  
\end{align} 
where $\k_{11}$ is the 11-dimensional gravitational constant, $T_2$ is the 
M2-brane tension,
\be T_2 = \left( \fracmm{2\p^2}{\k_{11}^2}\right)^{1/3}\lb{m2t} ~~,
\ee
$C$ is the 3-form gauge field of the eleven-dimensional supergravity \cite{Cremmer:1978km}, 
$F=d \wedge C$ is the four-form gauge field strength, $R$ is the gravitational scalar 
curvature in $D=11$, $\ve_{11}$ stands for the Levi-Civita symbol in eleven dimensions,
 while $(J,E_8,X_8)$ are the {\it quartic\/} 
polynomials with respect to the eleven-dimensional Riemann curvature. In particular, the $J_{11}$ is given by 
\be J_{11}=3\cdot 2^8\left(R^{mijn}R_{pijq}R\du{m}{rsp}R\ud{q}{rsn}
+\fracmm{1}{2}R^{mnij}R_{pqij}R\du{m}{rsp}R\ud{q}{rsn}\right)~,
\label{oco}
\ee
the $E_8$ can be written in terms of the Euler density in eight dimensions,
\be E_8= \fracmm{1}{3!}\ve^{abcm\low{1}n\low{1}\ldots m\low{4}n\low{4}}
\ve_{abc {m'}\low{1}{n'}\low{1}\ldots
{m'}\low{4}{n'}\low{4}}R\ud{{m'}\low{1}{n'}\low{1}}{m\low{1}n\low{1}}\cdots 
R\ud{{m'}\low{4}{n'}\low{4}}{m\low{4}n\low{4}} \label{euler}
\ee
and the $X_8$ is given by the gravitational 8-form 
\be
X_8 =\fracmm{1}{192\cdot (2\p^2)^4}\left[ \tr \hat{R}^4 - \fracmm{1}{4}(\tr \hat{R}^2)^2
\right]~~, \label{8form} \ee
where $\hat{R}$ stands for the spacetime curvature 2-form in eleven dimensions, and the traces are taken with respect to (implicit) Lorentz indices in $D=11$ dimensions. All (latin) vector indices take values $i,j,k,\ldots=0,1,2,\ldots,10$, while they are all suppressed in Eq.~(\ref{m11}) for simplicity. See Ref.~\cite{Eguchi:1980jx} about the notation of the exterior differential forms describing gravitational quantities in any dimension.

The $D=11$ gravity theory (\ref{m11}) can be compactified with the warp factor on the product
$S^3\times S^4$ of extra dimensions down to four spacetime dimensions, in the presence of fluxes needed for moduli stabilisation \cite{Douglas:2006es}.  Being interested in the gravitational sector of the effective field theory in $D=4$ dimensions, we can apply a simple dimensional reduction to the action (\ref{m11}) and ignore details of its compactification together with all moduli. Then only  the terms {\it quartic} in the full $D=4$ spacetime curvature  survive, while some of them may be represented as the Bel-Robinson tensor squared \cite{Iihoshi:2007vv},
\be
S_{\rm 4}  =  \fracmm{1}{2\k^2}\int d^{4}x\,\sqrt{-g}\left( R +\k^6\b T^2 \right)~,
\label{4a}
\ee
where all quantities are now in $D=4$ with $\k=1/M_{\rm Pl}$, and $\b$ is the new dimensionless coupling constant whose value is not determined from these simple considerations. 

The BR tensor is defined by \cite{Bel:1959uwe,Robinson:1959,Deser:1999jw}
\be
T^{iklm} \equiv R^{ipql}R\udu{k}{pq}{m} +{}^*R^{ipql}{}^*R\udu{k}{pq}{m} =R^{ipql}R\udu{k}{pq}{m} +R^{ipqm}R\udu{k}{pq}{l}-\ha g^{ik}R^{pqrl} R\du{pqr}{m}
\label{belr}
\ee
by analogy with the energy-momentum tensor of the Maxwell theory of electromagnetism,
\be T^{\rm Maxwell}_{ij}=F_{ik}F\du{j}{k}+{}^*F_{ik}{}^*F\du{j}{k}~~,\quad 
F_{ij}=\pa_iA_j-\pa_jA_i\lb{max}~~,
\ee
where the superscript $(*)$ means the dual tensor in $D=4$. As regards the curvature tensor, we have
\be {}^*R_{iklm}=\ha E_{ikpq}R\ud{pq}{lm}\lb{dcur}~~,
\ee
where $E_{iklm}=\sqrt{-g}\,\ve_{iklm}$ is Levi-Civita tensor. 

Then one finds \cite{Deser:1999jw,Iihoshi:2007vv}
\be
T^2_{ijkl}=  -\frac{1}{4}({}^*\!R_{ijkl}^{~2})^2 
+\frac{1}{4}({}^*R_{ijkl}R^{ijkl})^2 
 = \frac{1}{4}(P_4^2-E^2_4) =\frac{1}{4}(P_4+E_4)(P_4-E_4)~, \label{id2}
\ee
where we have also introduced the Euler and Pontryagin topological densities in $D=4$,
\be E_4= \fracm{1}{4}\ve_{ijkl}\ve^{mnpq}R\ud{ij}{mn}R\ud{kl}{pq}=
 {}^*R_{ijkl}{}^*R^{ijkl}
 \label{eu}
\ee
and
\be P_4= {}^*R_{ijkl}R^{ijkl}~~, \label{pont}
\ee
respectively. We use the book-keeping notation $A^2_{ijlk}\equiv A^{ijlk}A_{ijlk}$ for any rank-4 tensor $A$.

On-shell (by using the equations of motion in the EH gravity), one can show \cite{Bel:1959uwe,Robinson:1959}
that  the BR tensor is fully symmetric and traceless,
\be T_{ijkl}=T_{(ijkl)}~~,\qquad T^i_{ikl}=0~~, \label{brs}
\ee 
is covariantly conserved, 
\be \nabla^iT_{ijkl}=0~~, \label{brcon}
\ee
and has a {\it positive} "energy' density",
\be T_{0000}>0~~. \label{brpos}
\ee
The BR tensor is also related to the symmetric gravitational Landau-Lifshitz (LL) energy-momentum 
{\it pseudo}-tensor  \cite{Landau:1987gn} 
\begin{align}
(t_{LL})^{ij} = &  -\h^{ip}\h^{jq}\G^k_{pm}\G^m_{qk} 
+\G^i_{mn}\G^j_{pq}\h^{mp}\h^{nq}
-\left( \G^m_{np}\G^j_{mq}\h^{in}\h^{pq}
+\G^m_{np}\G^i_{mq}\h^{jn}\h^{pq}\right) \nonumber \\
&  +h^{ij}\G^{m}_{np}\G^n_{mq}\h^{pq}\label{pllt}
\end{align}
and the non-symmetric Einstein (E) gravitational energy-momentum pseudo-tensor 
 \cite{Held:1980gb} 
\be (t^E)^i_j = \left( -2\G^i_{mp}\G^m_{jq}+\d^i_j\G^n_{pm}\G^{m}_{qn}
\right)\h^{pq}
\label{epst}
\ee
in Riemann normal coordinates as follows \cite{Deser:1999jw}:
\be T_{ijkl}=\pa_k\pa_l \left( t^{LL}_{ij}
+\frac{1}{2}t^E_{ij}\right)~~.\label{elpsetr}
\ee

\section{Conclusion}

Having motivated the presence of the scalar curvature squared term from inflation and the Bel-Robinson-tensor squared term from M-theory, we propose the SBR gravity as a sum of them with the following action:
\begin{align} \label{sbr}
S_{\rm SBR}[g_{ij}] & = \fracmm{M^2_{\rm Pl}}{2}\int \mathrm{d}^4x\sqrt{-g} \left[
R +\fracmm{1}{6m^2}R^2 + \fracmm{\b}{8M^6_{\rm Pl} }T^2\right]~,\\
& = \fracmm{M^2_{\rm Pl}}{2}\int \mathrm{d}^4x\sqrt{-g} \left[ 
R +\fracmm{1}{6m^2}R^2 + \fracmm{\b}{32M^6_{\rm Pl} }(P^2_4 - E^2_4)\right]~,\nonumber
\end{align}
where we have used Eq.~(\ref{id2}).

Equation (\ref{sbr}) is our proposal for the $D=4$ gravitational effective action in the high-curvature regime, motivated by Starobinsky inflation and inspired by superstrings/M-theory. The high-curvature regime is defined by a situation where the values of
the second and third terms in Eq.~(\ref{sbr}) are comparable or larger than the value of the first one. Of course, the action (\ref{sbr}) is still an approximation that is not valid at the scales close to  the Planck scale.

The proposed action (\ref{sbr}) is gravitational (no matter added) and geometric, it does not have arbitrary functions and extra scalars (beyond spacetime metric and Starobinsky's scalaron that is the physical excitation of the higher-derivative $(R+\a R^2)$ gravity). 

The SBR action has only two parameters $(m,\b)$, which implies its predictive power for a wide range of physical applications such as black holes entropy \cite{Elgood:2020nls}, inflation \cite{Ivanov:2021chn} and Hawking radiation \cite{Ketov:1991jw}. These parameters should be subject to renormalization, i.e. they should depend upon a physical scale. Being subject to the renormalization of the parameters, the action (\ref{sbr}) may be applied at any scale.

The SBR action is superstring-inspired in the sense that its quartic terms in the curvature may be derivable from superstring/M-theory as the theory of quantum gravity that is not only renormalizable but is also unitary and ghost-free by construction \cite{Bachas:2008qjs}.  In other words, the popular Lovelock-like or Horndeski-like criteria demanding the absence of the higher-derivatives (beyond the 2nd order) in the equations of motion do not apply.

The Ricci-tensor-dependent terms in the gravitational effective action of superstring/M-theory cannot be determined from the known perturbative S-matrix in superstring theory, but they are not forbidden either \cite{Ketov:2000dy}. From the viewpoint of string theory, the coupling constants $(m,\b)$ are supposed to be given by the vacuum expectation values of moduli fields (including dilaton) as a result of compactification.
However, all those moduli and dilaton have to be stabilized, which is a difficult problem \cite{Douglas:2006es,Alexandrov:2016plh}.

The SBR action includes two topological densities (or 4-forms) $E_4$ and $P_4$ that can be locally represented as the wedge derivatives of the corresponding Chern-Simons 3-forms. However, they do contribute to the equations of motion in the SBR gravity because they enter the action (\ref{sbr}) as the squared terms. Since the Euler density $E_4$ is the same as the Gauss-Bonnet density  $G=R^2-4R^2_{ij} +R^{ijkl}R_{ijkl}$, when ignoring the $P^2_4$ term in Eq.~(\ref{sbr}), the
SBR action reduces to the particular modified gravity action of the type $F(R,G)$ with the quadratic terms in
$R$ and $G$ only, {\it cf.} Ref.~\cite{DeLaurentis:2015fea}). 

Specific applications of the proposed SBR theory will be studied elsewhere.

\section*{Acknowledgements}

The author was supported by Tokyo Metropolitan University, the Japanese Society for Promotion of Science under the grant No.~22K03624, the World Premier International Research Center Initiative (MEXT, Japan), and the Tomsk State University development program Priority-2030.

\bibliography{Bibliography}{}
\bibliographystyle{utphys}

\end{document}